\documentclass[final,,twoside]{IEEEtranTCOM}
\normalsize
\ifCLASSINFOpdf
\else
\fi

\usepackage{amsthm}
\usepackage{amsmath}
\usepackage{amssymb}
\usepackage{graphicx}
\usepackage{esint}
\usepackage{amsfonts}
\usepackage{cite}
\usepackage{balance}
\usepackage{caption}
\usepackage{subcaption}
\usepackage{epstopdf}
\usepackage{color}
\makeatletter

\theoremstyle{plain}
\newtheorem{thm}{\protect\theoremname}

\makeatother

\providecommand{\theoremname}{Theorem}

\theoremstyle{plain}

\makeatother

\providecommand{\lemmaname}{Lemma}

\DeclareMathOperator{\erf}{erf}

\begin{document}
\title{ Error analysis of mixed THz-RF wireless systems}

\author{
Alexandros-Apostolos A. Boulogeorgos, Senior Member, IEEE, and Angeliki Alexiou, Member, IEEE 

\thanks{The authors  are with the of Digital Systems, University of Piraeus Piraeus 18534 Greece (e-mails: al.boulogeorgos@ieee.org, alexiou@unipi.gr).
}

}
\maketitle	
\begin{abstract}
In this letter, we introduce a novel mixed terahertz (THz)-radio frequency (RF) wireless system architecture, which can be used for backhaul/fronthaul applications, and we deliver the theoretical framework for its performance assessment.
In more detail, after identifying the main design parameters and characteristics,  we derive novel closed-form expressions for the end-to-end signal-to-noise {ratio} cumulative density function, the outage probability and the symbol error rate, assuming that the system experiences  the joint effect of fading and stochastic antenna misalignment. 
The derived analytical framework is verified through simulations and quantifies the system's effectiveness and reliability. 
Finally, our results contribute to the extraction of useful design~guidelines.
\end{abstract}

\section{Introduction}\label{S:Intro}
Terahertz (THz) wireless systems have become a topic of much hype both in academia and industry, due to the high spectrum availability that they offer~\cite{A:THz_Technologies_to_deliver_optical_network_QoE_in_Wireless_Systems_beyond_5G,
WP:Wireless_Thz_system_architecture_for_networks_beyond_5G,C:UserAssociationInUltraDenseTHzNetworks}. 
In addition, they are license-free; hence, they are cost-effective relative to the radio frequency (RF) ones. 
Therefore, {THz wireless systems are expected to become} attractive backhaul solutions. 
On the other hand, they have considerably high sensitivity to blockage; especially, in urban {environment} with high obstacles density. Thus, their usage in fronthaul scenarios is questionable. This observation aspires  the investigation of mixed THz-radio frequency (RF) wireless systems as promising backhaul-fronthaul solutions. {This concept is envisioned to allow multiple RF {links} to feed one THz link.} Such a use case may be employed in harsh environments, where the fiber optic structure is under development in order to increase the installation flexibility and last mile quality of experience. 

Scanning the technical literature, a great amount of research effort has been put on modeling and evaluating the performance of both THz~\cite{A:Channel_modeling_and_capacity_analysis_for_electromagnetic_wireless_nanonetworks_in_the_THz_band,
C:Simplified_molecular_absorption_loss_model_for_275_400GHz_frequency_band,
C:Performance_Evaluation_of_THz_Wireless_Systems_Operating_in_475_400_GHz_band,
C:Frequency_domain_scattering_loss_in_THz_band,paper_toyrkoi,A:Channel_and_propagation_measurements_at_300_GHz,
A:Three_dimentional_end_to_end_modeling_and_analysis_for_graphene_enabled_THz_band_communications,
A:Analytical_Performance_Assessment_of_THz_Wireless_Systems} and RF wireless systems (see e.g., \cite{B:Alouini} and references therein). 
{In particular}, in~\cite{A:Channel_modeling_and_capacity_analysis_for_electromagnetic_wireless_nanonetworks_in_the_THz_band}, the authors reported a novel THz-band propagation model, whereas, in~\cite{C:Simplified_molecular_absorption_loss_model_for_275_400GHz_frequency_band}, a simplified molecular absorption loss model was delivered. 
The latter model was used in~\cite{C:Performance_Evaluation_of_THz_Wireless_Systems_Operating_in_475_400_GHz_band} for the quantification {of} the THz link capacity. 
The con of the above mentioned contributions is that they overestimated the THz system performance, since they neglect the impact of fading, which can be generated due to scattering on aerosols~\cite{C:Frequency_domain_scattering_loss_in_THz_band}. 
On the contrary, in~\cite{paper_toyrkoi} and~\cite{A:Channel_and_propagation_measurements_at_300_GHz}, the authors  modeled fading as stochastic processes. In particular, in~\cite{paper_toyrkoi}, they experimentally proved that the {envelope} of the fading coefficient follows Nakagami-m distribution under both non-line-of-sight and line-of-sight conditions.  Similarly, in~\cite{A:Channel_and_propagation_measurements_at_300_GHz}, the authors delivered experimental verifications of the existence of shadowing in the $300\text{ }\mathrm{GHz}$ band. Likewise, in~\cite{A:Three_dimentional_end_to_end_modeling_and_analysis_for_graphene_enabled_THz_band_communications}, the impact of antenna misalignment was investigated, while, in~\cite{A:Analytical_Performance_Assessment_of_THz_Wireless_Systems}, the joint impact of antenna misalignment and fading in THz wireless systems was evaluated in terms of outage probability (OP) and capacity. On the other hand, the performance of  relaying systems operating in the RF band have been studied in several works (see e.g.,\cite{Maged,ref13_Al_hard_imperf,A:Impact_of_IQI_on_AF_Relaying} and references therein). {The aforementioned} contributions assumed that the links at {both} hops {operate in the same frequency band; hence, the same fading distribution was assumed for both of them. However, this is not a valid assumption for links, which are established in different bands.} 

{To the best of the authors knowledge, despite their paramount importance, mixed THz-RF wireless systems have not been proposed and their performance has not been studied.} Motivated by this, in this letter, we introduce the mixed THz-RF wireless system architecture, we present the theoretical framework that quantifies its outage as well as error performance and provides useful design guidelines. 
In more detail, we report the system model that take into consideration the main design parameters as well as both the THz and RF channels {particularities}. In this direction, it is worth-noting that the fading coefficient {envelope} of the THz channel, in this work, is modeled as an $\alpha-\mu$ distribution. This distribution is configurable and can be simplified into Rayleigh, Nakagami-$m$ as well as Gamma in order to accommodate both the impact of multipath fading and shadowing. Moreover, in order to incorporate the effect of antenna misalignment, we modeled the elevation and horizontal displacement at the THz receiver plane as independent and identical Gaussian distributions. Building upon the system and channel models, we derive novel exact closed-form expressions for the cumulative density function (CDF) of the end-to-end (e2e) SNR, the OP and symbol error rate (SER). These expressions are verified through respective simulations and can be used to accurately evaluate the mixed THz-RF wireless system performance.   

\subsubsection*{Notations} 
The absolute value and exponential function are denoted by $|\cdot|$, and $\exp\left(x\right)$, respectively.
Likewise,  $\sqrt{x}$ returns the square root of $x$ and $\min\left(\cdot,\cdot\right)$ represented the minimum operator.  Moreover, $P_r\left(\mathcal{A}\right)$ is the probability that the event $\mathcal{A}$ is valid.  
The upper incomplete Gamma~\cite[eq. (8.350/2)]{B:Gra_Ryz_Book} and Gamma~\cite[eq. (8.310)]{B:Gra_Ryz_Book} functions are respectively denoted by $\Gamma\left(\cdot, \cdot\right)$, and $\Gamma\left(\cdot\right)$, while the Q-function is reprented by $Q\left(\cdot\right)$~\cite{B:Special_integral_functions_used_in_wireless_communications_theory}. Finally,  $G_{p, q}^{m, n}\left(x\left| \begin{array}{c} a_1, a_2, \cdots, a_{p} \\ b_{1}, b_2, \cdots, b_q\end{array}\right.\right)$  returns the Meijer G-function~\cite[eq. (9.301)]{B:Gra_Ryz_Book}, and $H_{p, q}^{m, n}\left[z \left|\begin{array}{c} (a_1, b_1), \cdots, (a_p, b_p) \\ (c_1, d_1), \cdots, (c_p, d_p) \end{array} \right. \right]$ stands for the Fox H-function~\cite[eq. (8.3.1/1)]{B:Prudnikov_v3}.

\section{System  model}\label{sec:SM}

We consider a mixed THz-RF dual-hop decode-and-forward (DF) system that is used for downlink. The direct link between the source (S) and destination (D) is assumed to be weak enough to be ignored. Moreover, a relay (R) is employed, which is equipped with a THz receiver and an RF transmitter.  
We assume that the S-R link is established in the THz band, whereas the R-D link is utilized in the RF~band. Likewise, it is assumed that R operates in half-duplex mode. Thus, in the first timeslot, R listens to S, whereas in the second timeslot, R, decodes, re-encodes and forwards to D the received signal. 
\subsection{THz link} 

The received signal at R can be expressed~as
\begin{align}
y_r = h_r s + w_r,
\end{align}
where $s$ and $w_r$ are respectively the {signal} transmitted by S and the zero-mean additive white Gaussian noise (AWGN),  with variance $N_{o,1}$. Additionally, $h_r$ represents the THz channel coefficient that can be analyzed~as
\begin{align}
h_r = h_l h_{pf}.
\label{Eq:h1}
\end{align} 
In~\eqref{Eq:h1}, $h_l$ is the deterministic path-gain and can be obtained, according to~\cite[eqs. (5)-(17)]{A:Analytical_Performance_Assessment_of_THz_Wireless_Systems}, {as 
\begin{align}
h_l = \frac{c \sqrt{G_{t,1} G_{r,1}}}{4\pi f_1 d_1} \exp\left(-\frac{1}{2}\kappa(f_1, T, \psi, p) d_1\right),
\end{align}
where $c$, $f_1$, and $d_1$ respectively denote the speed of light, the transmission frequency, and distance, while $G_{t,1}$, and  $G_{r,1}$ are the THz transmission and reception antenna gains. Likewise, $\kappa(f_1, T, \psi, p)$ is the molecular absorption coefficient, depends on the temperature, $T$, the relative humidity, $\psi$, as well as the atmospheric pressure, $p$, and can be calculated~as  
\begin{align}
\kappa(f_1, T, \psi, p) &= \frac{q_1 v (q_2 v + q_3)}{(q_4 v+q_5)^2 + \left(\frac{f_1}{100 c}-p_1\right)^2} 
\nonumber \\ &
+ \frac{q_6 v (q_7 v + q_8)}{(q_9 v + q_{10})^2 + \left(\frac{f_1}{100 c}-p_2\right)^2} 
\nonumber \\ &
+ c_1 f_1^3 + c_2 f_1^2 + c_3 f + c_4,
\end{align}
with $q_1 = 0.2205$, $q_2=0.1303$, $q_3=0.0294$, $q_4=0.4093$, $q_5=0.0925$, $q_6=2.014$, $q_7=0.1702$, $q_8=0.0303$, $q_9=0.537$, $q_{10}=0.0956$, $c_1 = 5.54\times 10^{-37} \text{ }\mathrm{Hz}^{-3}$, $c_2 = - 3.94\times 10^{-25} \text{ }\mathrm{Hz}^{-2}$, $c_3 = 9.06\times 10^{-14} \text{ }\mathrm{Hz}^{-1}$, $c_4 = - 6.36\times 10^{-3} \text{ }\mathrm{Hz}^{-3}$, $p_1=10.835\text{ }\mathrm{cm}^{-1}$, $p_2=12.664\text{ }\mathrm{cm}^{-1}$, 
$v=\frac{\psi}{100}\frac{p_w(T,p)}{p}$, and
 $p_w(T,p)$ is the saturated water vapor partial pressure in temperature T, which can be evaluated based {on} Buck's equation. }

Moreover, $h_{pf}$ is a random variable that accommodates the joint impact of fading and antenna misalignment, with probability density function (PDF) and CDF that can be respectively expressed as~{\cite[eqs. (26), and (27)]{A:Analytical_Performance_Assessment_of_THz_Wireless_Systems}}
\begin{align}
f_{|h_{fp}|}(x) = \phi 
S_0^{-\phi}  &
\frac{\mu^{\frac{\phi }{\alpha}}}{\hat{h}_f^{\alpha}\Gamma\left(\mu\right)} 
x^{\phi -1} 
\nonumber \\ & \times
\Gamma\left(
\frac{\alpha \mu - \phi }{\alpha}, 
\mu \frac{x^{\alpha}}{\hat{h}_f^{\alpha}} S_0^{-\alpha}
\right).
\label{Eq:f_{|h_{fp}|}}
\end{align}
and 
\begin{align}
F_{|h_{fp}|}(x) &= 1-\frac{1}{\alpha} \frac{x^{\phi }}{\hat{h}_{f}^{\phi }}  \frac{\phi }{S_0^{\phi }}  
\nonumber \\ & \times
\sum_{k=0}^{\mu-1} \frac{\mu^{\frac{\phi }{\alpha}}}{k!} 
\Gamma\left(\frac{\alpha k-\phi }{\alpha},{\mu\frac{x^{\alpha}}{\hat{h}_f^{\alpha}}} S_0^{-\alpha} \right),
\label{Eq:F_h_fp}
\end{align}   
where $\alpha > 0$ is the distribution parameter and $\mu$ is the normalized variance of the fading channel {envelope} that follows $\alpha-\mu$ distribution, $\hat{h}_f$ is the $\alpha$-root mean value of the fading channel {envelope}, $S_0$ is the fraction of the collected power when the transceivers antennas are fully-aligned  
{and can be evaluated~as
$S_0 = \left|\erf\left(\zeta\right)\right|^2,$
with $\zeta=\sqrt{\frac{\pi}{2}} \frac{r_1}{w_{d1}}$, $r_1$ and $w_{d1}$ respectively representing the radius of the reception antenna effective area and the transmission beam footprint radius at distance $d_1$.  
Additionally,} $\phi$ is the squared ratio of the equivalent beam width radius at R, $w_{e}$, to the doubled spatial jitter standard deviation, $\sigma_s$, {and can be defined~as
$\phi=\frac{w_e^2}{2\sigma_s^2}$, where $w_e^2=w_{d1}^2 \frac{\sqrt{\pi}\erf(\zeta)}{2\zeta\exp(-\zeta^2)}$.} 
 

\subsection{RF link}

The received signal at D is given~by
\begin{align}
y_2 = h_2 \tilde{s} + w_2,
\label{Eq:y2}
\end{align}   
where $\tilde{s}$, $h_2$ and $w_2$ denote respectively the R transmitted signal, the RF channel coefficient and the zero-mean AWGN with variance $N_{o,2}$. The channel coefficient of the RF link, $h_2$, can be expressed as
$h_2 = h_g h_f,$
where $h_g$ stands for the deterministic path-gain and can be obtained as {$h_g= \xi d_2^{-\eta_2/2}$}, with $\xi$,  $\eta_2$, and $d_2$  respectively being  the deterministic path-gain coefficient, the RF link path-loss exponent, and the R-D distance, while  $h_f$ denotes the fading channel coefficient. We assume that the {envelope}  of $h_f$ follows Rayleigh distribution. {Note that in urban environments, there are several objects that scatter the radio signal, before it arrives at the RX; hence, Rayleigh is a reasonable model for the RF link.}


\section{Performance analysis}
\subsection{End-to-end SNR statistics}

The e2e SNR of the mixed THz-RF wireless system can be expressed~as
\begin{align}
\gamma_e = \min\left(\gamma_{1}, \gamma_2\right),
\label{Eq:gamma_e}
\end{align} 
where
\begin{align}
\gamma_{1} = \frac{\left|h_l\right|^2 \left|h_{pf}\right|^2 E_{s}}{N_{o,1}}
\label{Eq:gamma_1}
\text{ and }
\gamma_{2} = \frac{\left|h_g\right|^2 \left|h_{f}\right|^2 E_{r}}{N_{o,2}},
\end{align} 
with $E_{s}$ and $E_{r}$ denoting the S and R transmission powers,~respectively. 

The following theorem returns an insightful closed-form expression for the CDF of the e2e SNR of the mixed THz-RF~link. 
\begin{thm}
The CDF of the e2e SNR of the mixed THz-RF~link can be obtained as in~\eqref{Eq:CDF_gamma_e_final}, given at the top of the next~page. 
\begin{figure*}
\vspace{-0.4cm}
\begin{align}
&F_{\gamma_e}\left(x\right) = 1 -  \frac{\phi }{\alpha} \left(\frac{ N_{o,1}}{S_0^2\hat{h}_{f}^{2}\left|h_l\right|^2 E_s} \right)^{\phi /2}     
\sum_{k=0}^{\mu-1} \frac{\mu^{\frac{\phi }{\alpha}}}{k!} x^{\phi /2} \exp{\left(- \frac{N_{o,2} x}{|h_g|^2 E_r}\right)}
\Gamma\left(\frac{\alpha k-\phi }{\alpha},{\mu \left(\frac{ N_{o,1}}{\left|h_l\right|^2 E_s} \right)^{\alpha/2} \frac{ x^{\alpha/2}}{\hat{h}_f^{\alpha} S_0^{\alpha}}}  \right)
 \label{Eq:CDF_gamma_e_final}
\end{align}\vspace{-0.5cm}
\hrulefill
\end{figure*}
\end{thm}
\vspace{-0.2cm}
\begin{IEEEproof}
Please refer to Appendix A.
\end{IEEEproof}

\textbf{Remark 1.}
{From~\eqref{Eq:CDF_gamma_e_final}}, the OP can be evaluated~as
$P_o\left(\gamma_{th}\right) =  F_{\gamma_e}\left(\gamma_{th}\right),$
where $\gamma_{th}$ is the SNR~threshold.
\subsection{Average SER}
The following theorem returns a closed-form expression for the average SER. 
\begin{thm}
The average SER of the mixed THz-RF wireless system can be analytically evaluated as in~\eqref{Eq:Average_P_e4}, given at the top of the next page. 
\begin{figure*} 
\vspace{-0.2cm} 
\begin{align}
\overline{P}_e &= \frac{a}{2}
- a \sqrt{\frac{b}{4\pi}} \frac{\phi  \mu^{\frac{\phi }{\alpha}}}{\alpha} \left(\frac{ N_{o,1}}{S_0^2\hat{h}_{f}^{2}\left|h_l\right|^2 E_s} \right)^{\phi /2}    \left(\frac{N_{o,2}}{|h_g|^2 E_r} + b\right)^{-\frac{\phi +1}{2}} 
\nonumber \\ & \times
\sum_{k=0}^{\mu-1} \frac{1}{k!} 
H_{3,3}^{2,2}\left(\frac{\mu \left(\frac{ N_{o,1}}{\left|h_l\right|^2 E_s} \right)^{\alpha/2} \frac{ 1}{\hat{h}_f^{\alpha} S_0^{\alpha}}}{\left(\frac{N_{o,2}}{|h_g|^2 E_r} + b\right)^{\alpha/2}} \left|
\begin{array}{c} 
\left(-\frac{\phi +1}{2}, \frac{\alpha}{2}\right), \left(\frac{1-\phi }{2}, \frac{\alpha}{2}\right), \left(1, 1\right) \\ 
\left( \frac{\alpha k-\phi }{\alpha}, 1\right), \left(0, 1\right), \left(-\frac{\phi +1}{2}, \frac{\alpha}{2}\right) 
\end{array}
 \right.\right)
\label{Eq:Average_P_e4}
\end{align} \vspace{-0.4cm}
\hrulefill
\end{figure*}
In~\eqref{Eq:Average_P_e4},  $a$ and $b$ are modulation-specific~constants ({e.g., binary shift keying (BPSK): $a=1$ and $b=0.5$, quadrature phase shift keying (QPSK) $a=1$ and $b=0.25$, and $M-$ quadrature amplitude modulation (QAM): $a=4$ and $b=\frac{3}{M-1}$}).
\end{thm}
\begin{IEEEproof}
Please refer to Appendix B.
\end{IEEEproof}

\section{Results \& Discussion}
In this section, we illustrate the outage and error performance of the mixed THz-RF wireless system by providing analytical and  simulation results for different design parameters. In particular, we consider the following insightful scenario. Unless otherwise stated, it is assumed that   $\alpha=1$, $\mu=2$, while the relative humidity, the atmospheric pressure and temperature are respectively $50\text{ }\%$, $101325\text{ }\mathrm{Pa}$, and $296\text{ }^oK$. Moreover, the operation frequency of the THz link is  $275\text{ }\mathrm{THz}$, while the THz antenna gains in both S and R equal $55\text{ }\mathrm{dBi}$. Moreover, the spatial jitter standard deviation, $\sigma_s$, is assumed to be equal to $10\text{ }\mathrm{mm}$. Likewise,  the S-R transmission distance is assumed to be equal to $20\text{ }\mathrm{m}$, {while, for the R-D link, $\eta_2=2$, and $\xi$ can be obtained as $\xi=\frac{c G_t^r G_r^r}{4\pi f_r}$, where $f_r$, $G_t^r$ and $G_r^r$ are respectively the transmission frequency of the RF link, which is set to $800\text{ }\mathrm{MHz}$, the transmission and reception antenna gains at S and R. Finally, it is assumed that $\frac{G_t^r G_r^t}{d_2^2}=1\text{ }\mathrm{m}^{-2}$.}  In the following figures, the numerical results are shown with continuous or/and dashed lines, while markers are employed to illustrate the simulation results.  

\begin{figure}
\centering\includegraphics[width=0.8\linewidth,trim=0 0 0 0,clip=false]{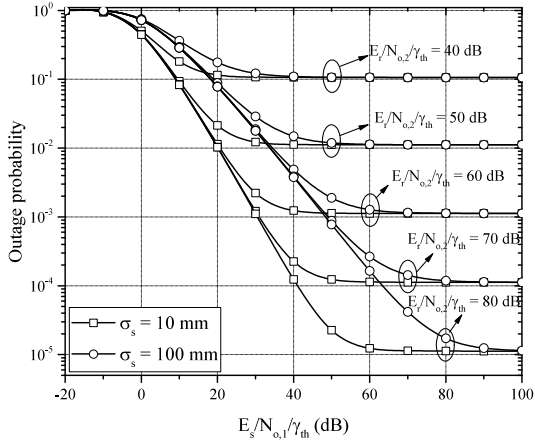}\vspace{-0.1cm}
\caption{
OP vs $\frac{E_s/N_{o,1}}{\gamma_{th}}$, for different {values} of $\frac{E_r/N_{o,2}}{\gamma_{th}}$ and $\sigma_s$.
}
\label{fig:OP}
\end{figure}  

Fig.~\ref{fig:OP} presents the outage performance of the mixed THz-RF wireless system as a function of $\frac{E_s/N_{o,1}}{\gamma_{th}}$, for different {values} of $\frac{E_r/N_{o,1}}{\gamma_{th}}$ and $\sigma_s$. As expected, for a given $\frac{E_r/N_{o,2}}{\gamma_{th}}$ and $\sigma_s$, as $\frac{E_s/N_{o,1}}{\gamma_{th}}$ {increases}, the OP decreases. Moreover, for a fixed $\frac{E_s/N_{o,1}}{\gamma_{th}}$ and $\sigma_s$, as $\frac{E_r/N_{o,2}}{\gamma_{th}}$, the outage performance improves. Interestingly, the outage performance are constrained by {the worst link}. Finally, from this figure, it is seen that, for a given $\frac{E_s/N_{o,1}}{\gamma_{th}}$ and $\frac{E_r/N_{o,2}}{\gamma_{th}}$, as $\sigma_s$ increases, the outage performance degrades. This indicates the importance of taking into consideration the impact of antenna misalignment in the performance assessment of the mixed THz-RF wireless~system.   

\begin{figure}
\centering
\includegraphics[width=0.8\linewidth,trim=0 0 0 0,clip=false]{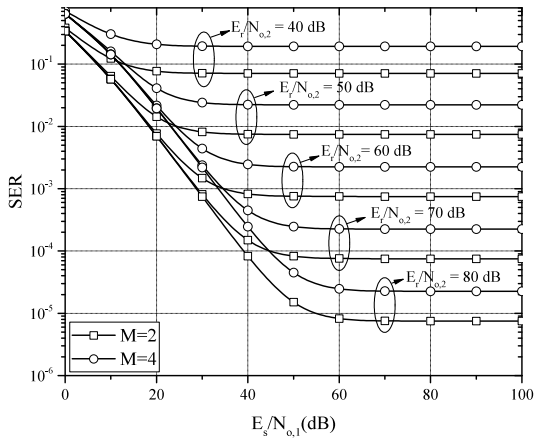}\vspace{-0.1cm}
\caption{
SER vs ${E_s/N_{o,1}}$, for different {values} of ${E_r/N_{o,2}}$ and~$M$.
}
\label{fig:SER1}
\end{figure} 

Fig.~\ref{fig:SER1}, the SER of a mixed THz-RF wireless system, which employs $M-$QAM, as a function of ${E_s/N_{o,1}}$, for different values of ${E_r/N_{o,2}}$ and~$M$ is plotted. {Note that $M=2$ corresponds to the SER performance of BPSK.} We observe that, for a given ${E_r/N_{o,2}}$ and $M$, as ${E_s/N_{o,1}}$ increases, the error performance improves. Similarly, for a fixed  ${E_s/N_{o,1}}$ and $M$,  as ${E_r/N_{o,2}}$ increases, the SER decreases. Meanwhile, we observe that the minimum SER is determined by the transmission SNR  of the worst link.  Finally, for given ${E_s/N_{o,1}}$ and ${E_r/N_{o,2}}$, as $M$ increases, the error performance degrades. This indicates that in order to improve the error performance of the mixed THz-RF wireless system, we can either increase the transmission SNR of the worst link, or decrease $M$.

\begin{figure}
\centering\includegraphics[width=0.8\linewidth,trim=0 0 0 0,clip=false]{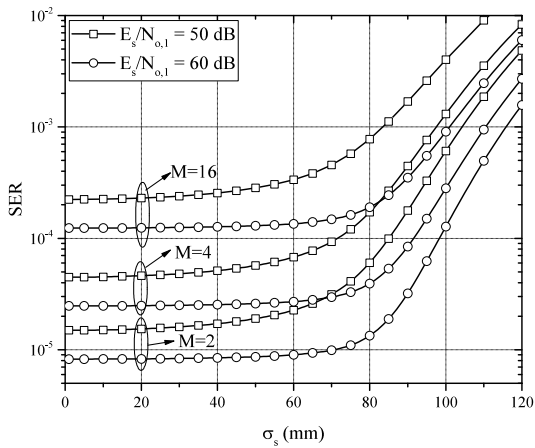}\vspace{-0.1cm}
\caption{
SER vs $\sigma_s$, for different {values} of ${E_s/N_{o,1}}$ and~$M$.
}
\label{fig:SER2}
\end{figure} 

Fig.~\ref{fig:SER2}, the SER of a mixed THz-RF wireless system, which employs $M-$QAM, as a function of $\sigma_s$, for different values of ${E_s/N_{o,1}}$ and~$M$, assuming ${E_r/N_{o,2}}=80\text{ }\mathrm{dB}$, is depicted. From this figure, we observe that, for fixed ${E_s/N_{o,1}}$ and $M$, as $\sigma_s$ increases, the error performance degrades. Moreover, for given $\sigma_s$ and $M$, as ${E_s/N_{o,1}}$ increases, the error performance improves, whereas, for fixed  $\sigma_s$ and ${E_s/N_{o,1}}$, as $M$ increases, the SER also increases. Finally, it is noticeable that in the relatively low $\sigma_s$ regime, the decrease of $M$ causes a more significantly boost in the error performance of the mixed THz-RF wireless system compared to the one that is caused by the increase of ${E_s/N_{o,1}}$. The reverse happens in the high $\sigma_s$ regime. This reveals the importance of taking into account the impact of antenna misalignment in the design of adaptive modulation and power allocation algorithms. 
\section{Conclusions}
This letter introduced the mixed THz-RF wireless system architecture and provided its outage and error performance assessment study. {Specifically}, novel closed-form expressions for the OP and SER were extracted, which accurately quantify the system performance and takes into consideration the transceivers characteristics and both the THz and RF channel particularities. Our results revealed that the performance of the mixed THz-RF wireless system are constrained  from the quality of the worst link as well as the importance of taking into account the impact of the THz antenna misalignment, when assessing the system~reliability. 
 \section*{Acknowledgment}
 This work has received funding from the European Commission’s Horizon 2020 research and innovation programme under grant agreement No. 761794 (TERRANOVA).
 The authors  would like to thank the editor and the anonymous reviewers for their constructive comments and criticism.
\section*{Appendix}
\section*{Appendix A}
Based on~\eqref{Eq:gamma_e}, the CDF of $\gamma_e$ can be obtained~as
\begin{align}
F_{\gamma_e}\left(x\right) = F_{\gamma_1}\left(x\right) + F_{\gamma_2}\left(x\right) - F_{\gamma_1}\left(x\right) F_{\gamma_2}\left(x\right), 
\label{Eq:CDF_gamma_e}
\end{align}
where $F_{\gamma_1}\left(\cdot\right)$ and $F_{\gamma_2}\left(\cdot\right)$ are respectively the CDFs of $\gamma_1$ and~$\gamma_2$. Next, we provide closed form expressions for $F_{\gamma_1}\left(\cdot\right)$ and $F_{\gamma_2}\left(\cdot\right)$.

The CDF of $\gamma_1$ can be analytically evaluated~as
$ F_{\gamma_1}\left(x\right) = P_{r}\left(\gamma_1\leq x\right), $
which, according to~\eqref{Eq:gamma_1}, can be rewritten~as
\begin{align}
 F_{\gamma_1}\left(x\right) = F_{\left|h_{pf}\right|}\left(\sqrt{\frac{x N_{o,1}}{\left|h_l\right|^2 E_s}}\right),
 \label{Eq:CDF_gamma_1s2}
\end{align}
or equivalently
\begin{align}
 & F_{\gamma_1}\left(x\right) =1-\frac{1}{\alpha} \left(\frac{ N_{o,1}}{\left|h_l\right|^2 E_s} \right)^{\phi /2} \frac{x^{\phi /2}}{\hat{h}_{f}^{\phi }}  \frac{\phi }{S_0^{\phi }}  
\nonumber \\ & \times
\sum_{k=0}^{\mu-1} \frac{\mu^{\frac{\phi }{\alpha}}}{k!} 
\Gamma\left(\frac{\alpha k-\phi }{\alpha},{\mu \left(\frac{ N_{o,1}}{\left|h_l\right|^2 E_s} \right)^{\alpha/2} \frac{ x^{\alpha/2}}{\hat{h}_f^{\alpha} S_0^{\alpha}}}  \right).
\label{Eq:CDF_gamma_1}
\end{align}

Similarly, since $|h_f|$ follows a Rayleigh distribution, $|h_f|^2$ follows an exponential distribution; hence, $\gamma_{2}$ also follows an exponential distribution with CDF that can be obtained~as
\begin{align}
 F_{\gamma_2}\left(x\right) =1-\exp{\left(- \frac{N_{o,2} x}{|h_g|^2 E_r}\right)}.
 \label{Eq:CDF_gamma_2}
\end{align}

By substituting~\eqref{Eq:CDF_gamma_1} and~\eqref{Eq:CDF_gamma_2} into~\eqref{Eq:CDF_gamma_e} and after some algebraic manipulations, we get~\eqref{Eq:CDF_gamma_e_final}. This concludes the proof.

\section*{Appendix B}
By assuming two-dimensional modulation, the conditional SER can be obtained as~\cite{B:Digital_Communications}
\begin{align}
P_e(x) = a \mathrm{Q}\left(\sqrt{2 b x}\right).
\end{align}
{Note that for several Gray bit-mapped constellations, the conditional SER can be expressed in this form~\cite{5722051,5397898}.}
Thus, the average SER can be expressed~as
\begin{align}
\overline{P}_e = \int_{0}^{\infty}P_e\left(x\right) f_{\gamma_{e}}(x) \mathrm{dx},
\label{Eq:Average_P_e}
\end{align}
or equivalently
\begin{align}
\overline{P}_e = - \int_{0}^{\infty} F_{\gamma_{e}}(x) f_{e}(x) \mathrm{dx}
\label{Eq:Average_P_e2}
\end{align}
{where $f_{{e}}(x)$ can be evaluated~as}
$f_{e}(x) = \frac{\mathrm{d}P_e\left(x\right)}{\mathrm{dx}},$
or
\begin{align}
f_{e}(x) = -a \sqrt{\frac{b}{4\pi}} x^{-1/2} \exp\left(-b x\right). 
\label{Eq:f_e_2}
\end{align}  

By substituting~\eqref{Eq:CDF_gamma_e_final} and~\eqref{Eq:f_e_2} into~\eqref{Eq:Average_P_e2}, we can  {write~\eqref{Eq:Average_P_e2}~as}
\begin{align}
\overline{P}_e &= \mathcal{I}_1 
- a \sqrt{\frac{b}{4\pi}} \frac{\phi }{\alpha} \left(\frac{ N_{o,1}}{S_0^2\hat{h}_{f}^{2}\left|h_l\right|^2 E_s} \right)^{\phi /2}     
\sum_{k=0}^{\mu-1} \frac{\mu^{\frac{\phi }{\alpha}}}{k!} \mathcal{I}_2(k),
\label{Eq:Average_P_e3}
\end{align} 
where
\begin{align}
\mathcal{I}_1 =a \sqrt{\frac{b}{4\pi}}\int_{0}^{\infty} x^{-1/2} \exp\left(-b x\right) \mathrm{dx}
\label{Eq:I1}
\end{align}
and
\begin{align}
\mathcal{I}_2(k) &= \int_{0}^{\infty}  x^{\left(\phi -1\right)/2} \exp{\left(- \left(\frac{N_{o,2} }{|h_g|^2 E_r}+b\right) x\right)}
\nonumber \\ & \times
\Gamma\left(\frac{\alpha k-\phi }{\alpha},{\mu \left(\frac{ N_{o,1}}{\left|h_l\right|^2 E_s} \right)^{\alpha/2} \frac{ x^{\alpha/2}}{\hat{h}_f^{\alpha} S_0^{\alpha}}}  \right)   \mathrm{dx}.
\label{Eq:I2}
\end{align}
By employing~\cite[eq. (3.326/2)]{B:Gra_Ryz_Book},~\eqref{Eq:I1} can be analytically evaluated~as
\begin{align}
\mathcal{I}_1 = \frac{a}{2},
\label{E	q:I1_final}
\end{align}
while by using~\cite[eq. (8.352/2)]{B:Gra_Ryz_Book},~\eqref{Eq:I2} can be rewritten~as
\begin{align}
\mathcal{I}_2(k) &= \int_{0}^{\infty}  x^{\left(\phi -1\right)/2} \Gamma\left(1, \left(\frac{N_{o,2} }{|h_g|^2 E_r}+b\right) x\right)
\nonumber \\ & \times
\Gamma\left(\frac{\alpha k-\phi }{\alpha},{\mu \left(\frac{ N_{o,1}}{\left|h_l\right|^2 E_s} \right)^{\alpha/2} \frac{ x^{\alpha/2}}{\hat{h}_f^{\alpha} S_0^{\alpha}}}  \right)   \mathrm{dx}.
\label{Eq:I2_s1}
\end{align}
Moreover, by employing~\cite[eq. (5)]{A:Exact_evaluations_of_some_Meijer_G_functions}, \eqref{Eq:I2_s1} can be obtained~as
\begin{align}
&\mathcal{I}_2(k) = \int_{0}^{\infty}  x^{\left(\phi -1\right)/2} 
G_{1,2}^{2,0}\left( \left(\frac{N_{o,2} }{|h_g|^2 E_r}+b\right) x \left| \begin{array}{c} 1 \\ 1, 0\end{array}\right.\right)
\nonumber \\ \hspace{-0.2cm} & \times\hspace{-0.1cm}
G_{1,2}^{2,0}\left(  \left(\frac{ N_{o,1}}{\left|h_l\right|^2 E_s} \right)^{\alpha/2} \hspace{-0.2cm} \frac{ \mu}{\hat{h}_f^{\alpha} S_0^{\alpha}} x^{\alpha/2} \left| \begin{array}{c} 1 \\ \frac{\alpha k-\phi }{\alpha},0 \end{array}\right.\right) \mathrm{dx},
\end{align}
which, based on~\cite[Ch. 2.3]{B:The_H_function}, can be written in closed-form~as 
\begin{align}
&\mathcal{I}_2(k) = \left(\frac{N_{o,2}}{|h_g|^2 E_r} + b\right)^{-\frac{\phi +1}{2}} 
\nonumber
\\ &\hspace{-0.1cm} \times \hspace{-0.1cm}
H_{3,3}^{2,2}\hspace{-0.1cm}\left(\hspace{-0.1cm}\frac{\mu \left(\frac{ N_{o,1}}{\left|h_l\right|^2 E_s} \right)^{\alpha/2} \hspace{-0.1cm} \frac{ 1}{\hat{h}_f^{\alpha} S_0^{\alpha}}}{\left(\frac{N_{o,2}}{|h_g|^2 E_r} + b\right)^{\alpha/2}} \hspace{-0.1cm} \left|\hspace{-0.2cm}
\begin{array}{c} 
\left(-\frac{\phi +1}{2}, \frac{\alpha}{2}\right), \left(\frac{1-\phi}{2}, \frac{\alpha}{2}\right), \left(1, 1\right) \\ 
\left( \frac{\alpha k-\phi}{\alpha}, 1\right), \left(0, 1\right), \left(-\frac{\phi +1}{2}, \frac{\alpha}{2}\right) 
\end{array} \hspace{-0.3cm}
 \right.\right)\hspace{-0.1cm}.
 \label{Eq:I2_s3}
\end{align}
Finally, by substituting~\eqref{Eq:I1} and~\eqref{Eq:I2_s3} into~\eqref{Eq:Average_P_e3}, we obtain~\eqref{Eq:Average_P_e4}. This concludes the proof. 

\balance
\bibliographystyle{IEEEtran}
\bibliography{IEEEabrv,References}

\end{document}